\pdfoutput=1
\documentclass[a4paper,fleqn,usenatbib]{mnras}

\def\nhu{\textit{$\rm cm^{-2}$}}
\def\nh{\textit{$N_{\rm H}$}}

\def\ergs{\textit{$\rm erg\ s^{-1}$}}

\def\um{\textit{$\rm \mu m$}}

\def\civ{\text{C\ \sc{iv}}}
\def\ciii{\text{C\ \sc{iii]}}}
\def\siv{\text{Si\ \sc{iv}}}
\def\nv{\text{N\ \sc{v}}}
\def\pv{\text{P\ \sc{v}}}
\def\lya{\textit{$\rm Ly \sc{\alpha}$}}

\def\cm{\textit{$\rm 10^{23}\ cm^{-2}$}}

\def\>{\textit{$\textgreater$}}
\def\<{\textit{$\textless$}}

\def\vmax{\textit{$V_{\rm max}$}}

\def\edd{\textit{$\rm \lambda_{Edd}$}}

\def\lbol{\textit{$L_{\rm bol}$}}

\def\fuse{\textit{FUSE}}
\def\chandra{\textit{Chandra}}

\def\wise{\textit{WISE}}
\def\swift{\textit{Swift}}
\def\hst{\textit{HST}}
\def\uu{\textit{U}}
\def\vv{\textit{V}}
\def\bb{\textit{B}}
\def\uvw{\textit{UVW}1}
\def\uvww{\textit{UVW}2}
\def\uvm{\textit{UVM}2}
\def\wpvs{$\rm WPVS\ 007$}

\def\cm{\textit{$\rm 10^{23}\ cm^{-2}$}}

\usepackage{newtxtext,newtxmath}
\usepackage[T1]{fontenc}
\usepackage{ae,aecompl}
\usepackage{graphicx}	
\usepackage{amsmath}	
\usepackage{amssymb}	

\title[Variability of WPVS 007]{On the Origin of the Dramatic Spectral Variability of WPVS 007}

\author[Li et al.]{
Junyao Li,$^{1,2}$\thanks{E-mail: lijunyao@mail.ustc.edu.cn}
Mouyuan Sun,$^{1,2}$\thanks{E-mail: ericsun@ustc.edu.cn}
Tinggui Wang,$^{1,2}$
Zhicheng He$^{1,2}$
and Yongquan Xue$^{1,2}$\thanks{E-mail: xuey@ustc.edu.cn}
\\\\
$^{1}$CAS Key Laboratory for Research in Galaxies and Cosmology, Department of Astronomy, University of Science and Technology of China, Hefei 230026, China\\
$^{2}$School of Astronomy and Space Science, University of Science and Technology of China, Hefei 230026, China
}

\date{Accepted XXX. Received YYY; in original form ZZZ}

\pubyear{2019}

\begin{document}
\label{firstpage}
\pagerange{\pageref{firstpage}--\pageref{lastpage}}
\maketitle

\begin{abstract}
\noindent We report the discovery of large-amplitude mid-infrared variabilities (MIR; $\sim 0.3$ mag) in the {\it{Wide-field Infrared Survey Explorer}} $W1$ and $W2$ bands
of the low-luminosity narrow-line Seyfert 1 galaxy WPVS 007, which exhibits prominent and varying broad-absorption lines (BALs) with blueshifted velocity up to $\sim 14000$ $\rm km\ s^{-1}$. The observed significant MIR variability, the UV to optical color variabilities in the \swift~bands that deviate from the predictions of pure dust attenuation models, and the fact that \swift~light curves can be well fitted by the stochastic AGN variability model suggest that its observed flux variabilities in UV-optical-MIR bands should be intrinsic, rather than owing to variable dust extinction. 
Furthermore, the variations of BAL features (i.e., trough strength and maximum velocity) and continuum luminosity are concordant. Therefore, we propose that the BAL variability observed in WPVS 007 is likely induced by the intrinsic ionizing continuum variation, alternative to the rotating-torus model proposed in a previous work. The BAL gas in \wpvs~might be in the low-ionization state as traced by its weak \nv~BAL feature; as the ionizing continuum strengthens, the \civ~and \siv~column densities also increase, resulting in stronger BALs and the emergence of high-velocity components of the outflow. The outflow launch radius might be as small as $\sim 8 \times 10^{-4}$ pc under the assumption of being radiatively driven, but a large-scale origin (e.g., torus) cannot be fully excluded because of the unknown  effects from additional factors, e.g., the magnetic field.
\end{abstract}


\section{Introduction}

Outflows are commonly observed in active galactic nuclei \citep[AGNs; e.g.,][]{Tombesi2010, Gofford2013, Harrison2014}, which are possibly driven by radiation pressure \citep[e.g.,][]{Proga2000, Proga2004}, magnetic field
\citep[e.g.,][]{Ohsuga2009, Jiang2014}, or Compton heating \citep[e.g.,][]{Begelman1983} from the accretion disk. Powerful outflows can be identified via blueshifted broad-absorption lines (BALs; FWHM \> 2000 $\rm km\ s^{-1}$) in either UV or X-ray bands, and are believed as an important candidate for AGN feedback by injecting their kinetic energy and momentum outward (see, e.g., \citealt{Fabian2012} and \citealt{King2015} for reviews). However, direct measurements of outflow geometry, distance, stability, energy and hence its influence on galaxy environment and evolution are so far inconclusive. Multi-epoch observations and variability studies that focus on sources with prominent outflow features provide a new way to constrain these properties, and thus help shed light on how and where the winds are launched, accelerated, varied, and coupled with galactic-wide materials \citep[e.g.,][]{FilizAk2012, Grier2015, Rogerson2016, Matzeu2017, Parker2017,  McGraw2018, Pinto2018, DeCicco2018}.

Previous studies have found that two mechanisms are mainly responsible for the observed BAL variability: (1) changes in the ionization state; and (2) changes of the covering factor (e.g., the absorbers moving into/out-of the sightline). The first mechanism can easily explain the coordinated relationships between BALs and continuum variabilities found in large quasar samples \citep[e.g.,][]{Wang2015, He2017}; and the variability timescale, which provides important information of the recombination timescale, can be used to constrain crucial outflow properties such as gas density, outflow distance and mass outflow rate \citep[e.g.,][]{Gofford2014, He2019}. While in the second scenario, the cloud transverse motion (thus BAL variability) is expected to be independent of the continuum, and the variability (transverse) timescale can be used to constrain the distance of the absorber \citep[e.g.,][]{Capellupo2012, Matzeu2016, Rogerson2016, Braito2018, Matzeu2019}. Therefore, distinguishing different variability mechanisms is crucial for us to investigate the physical conditions of the outflow. 

The low-luminosity ($M_V \approx -19.7$) narrow-line Seyfert~1 galaxy (NLS1) WPVS 007, located at $z = 0.02882$ \citep{Grupe1995}, is one of the most extensively studied source owing to its peculiar variability behavior \citep[e.g.,][]{Leighly2009, Grupe2013, Leighly2015}. 
At its first discovery in the \textit{ROSAT} all-sky survey, it appeared as a bright X-ray source with an extremely soft spectrum \citep{Grupe1995}. However, in the subsequent observations by \chandra~\citep{Vaughan2004} and \swift~observatories \citep{Grupe2007}, it showed persistently weak X-ray emission with no detections in most of the time, while the first hard X-ray detection suggested the existence of a high column-density ($\nh \sim 10^{22-23}$~cm$^{-2}$), partial-covering absorber \citep{Grupe2008}.

The spectroscopic monitoring by the \textit{Hubble Space Telescope} (\hst) and the \textit{Far Ultraviolet Spectroscopic Explorer} (\fuse) also revealed some unique phenomena. In the 1996 \hst~observation, WPVS 007 only exhibited a mini-BAL with a maximum velocity $V_{\rm max} \sim 900$~$\rm km\ s^{-1}$ \citep{Leighly2009}. However, a significant outflow feature traced by the blueshifted \pv~BAL with $V_{\rm max} \sim 6000$~$\rm km\ s^{-1}$ appeared in the 2003 \fuse~spectrum \citep{Leighly2009}, and a more dramatic \civ~BAL with $V_{\rm max} \sim 13000$~$\rm km\ s^{-1}$ was detected in the 2010 \hst~observation \citep{Leighly2015}.  The discovery of these prominent outflow features makes the low-luminosity \wpvs~a distinct object since powerful outflows are more likely to be found in luminous quasars \citep[e.g., PDS 456; ][]{Pounds2003, Reeves2003, Hamann2008, Dunn2010, Danehkar2018}. 
Interestingly, in the three succeeding \hst~observations from 2013 to 2015, the wind velocities decreased persistently down to a few thousand $\rm km\ s^{-1}$, accompanied with dimming and reddening of the continuum \citep{Leighly2015}. The significant velocity ``shifts'' (\citealt{Leighly2015}; also see our Section \ref{subsec:velocity_lumin}) make WPVS 007 an extraordinary object owing to the fact that most of observed BAL troughs only show variations in profile and strength rather than velocity structure \citep[e.g.,][]{Grier2016}. 

The coordinated correlation between outflow velocity and continuum flux is usually unexpected in the moving cloud scenario. However, \cite{Leighly2015} (hereafter L15) proposed an interesting scenario that outflows are launched from an inhomogeneous rotating torus to explain the coordination. As the torus rotates, the scale height varies, the extinction changes, and outflows with different launching velocities move into our sightline, which result in both changes on continuum fluxes and outflow velocities, thus building the bridge between velocity and continuum variabilities (see their Figure 5 and Section 3.4). 
However, we note that the BAL strength is also correlated with continuum flux (see Section \ref{subsec:ionization}), but the physical reason behind such a relationship has not been explored yet.

In addition, L15 intensively analyzed the UV-optical photometries for \wpvs~and captured a particular ``occultation event'' which lasted for $\sim60$ days in 2015. By assuming Keplerian motion of the occulting cloud, they constrained the distance of the absorber to be at the torus scale, and explained the occultation as a result of variable reddening in the rotating torus framework (see their Figure 1 and Figure 5 and our Section 4.4). If variable reddening also governs the flux variability during the $\sim$ten-years \swift~monitoring \citep{Grupe2013}, we should expect no significant mid-infrared (MIR) variability since the reprocessed MIR emission is insensitive to dust extinction. 

However, we check the \textit{Wide-field Infrared Survey Explorer} (\wise) data for this source and find large-amplitude variabilities ($\sim 0.3\ \rm mag$) in $W1$ ($3.4\ \um$) and $W2$ ($4.6\ \um$) bands, suggesting that \wpvs~must experience significant intrinsic luminosity variability. Although L15 can explain the occultation event in the variable extinction scenario,
how the additional variable factor (i.e., the intrinsic luminosity) would influence our understanding of the dramatic BAL variability in this particular source is currently unclear. Therefore, in this work, we plan to extensively analyze the combined UV-optical-IR variabilities of WPVS 007 aiming at comprehensively exploring its long-term variability nature and the driving mechanism of its coordinated BAL variabilities, which might give important clues on constraining physical properties of this unique low-luminosity, NLS1-BAL system. This paper is organized as follows. In $\S$ \ref{sec:observation} we describe the observations and data reduction. In $\S$ \ref{sec:variability} we present our analyses on \wise~variabilities, color variabilities and \swift~light curves that lead us to the conclusion that the variable intrinsic luminosity is mainly responsible for the flux variabilities observed in WPVS 007, instead of variable extinction. In $\S$ \ref{sec:discussion} we discuss the origin of its dramatic BAL variability in the variable luminosity framework and put important constraints on the driving mechanism and launch radius of the varying outflows. The results are summarized in $\S$ \ref{sec:conclusion}. Throughout this paper, we adopt a flat cosmology with $\rm H_0 = 70.0\ km\ s^{-1}\ Mpc^{-1}$, $\rm \Omega_M = 0.30$, and $\rm \Omega_\Lambda = 0.70$. 

\section{Observations and data reduction}
\label{sec:observation}

The archival \wise~MIR photometries, \swift~UV-optical photometries and \hst~UV-optical spectra used in this work are summarized in Table \ref{table:log}. Below we give a detailed description of each data set.

\begin{table*}
\centering
\caption{Information of archival \wise, \swift~and 
\hst~data used in this work}
\begin{tabular}{c c c c c c c} 
\hline
\hline
Telescope / Instrument & Observation Date & Photometric Bands / Spectral Coverages & PI\\
\hline
\textit{WISE}  & 2010 May  -- 2016 Nov  & {\it{W1}} (3.6~\um) and {\it{W2}} (4.5~\um) & {\it WISE}\\

\swift~/ XRT & 2005 Oct  -- 2018 Mar  & \uvww, \uvw, \uvm, \uu, \bb, \vv & {\it Swift} \& Dirk Grupe\\

\hst~/ FOS & 1996 Jul & 1140~\AA -- 2508~\AA~(BL) and 1573~\AA -- 6872~\AA~(RD) & Robert Goodrich\\

\hst~/ COS & 2010 Jun & 1230~\AA -- 2050~\AA~(FUV) and 1700~\AA -- 3200~\AA~(NUV) & Karen Leighly\\

\hst~/ COS & 2013 Jun & 1230~\AA -- 2050~\AA~(FUV) and 1700~\AA -- 3200~\AA~(NUV) & Karen Leighly\\ 

\hst~/ COS & 2013 Dec & 1230~\AA -- 2050~\AA~(FUV) and 1700~\AA -- 3200~\AA~(NUV) & Karen Leighly\\

\hst~/ COS & 2015 Mar & 1230~\AA -- 2050~\AA~(FUV) and 1700~\AA -- 3200~\AA~(NUV) & Karen Leighly\\

\hst~/ COS & 2017 Mar & 1230~\AA -- 2050~\AA~(FUV) and 1700~\AA -- 3200~\AA~(NUV) & Karen Leighly\\

\hline
\end{tabular}

\vspace{10.0 pt}
\label{table:log}
\end{table*}

\subsection{\textit{WISE} observations}  
\begin{figure*}
\centering
\includegraphics[width=\linewidth]{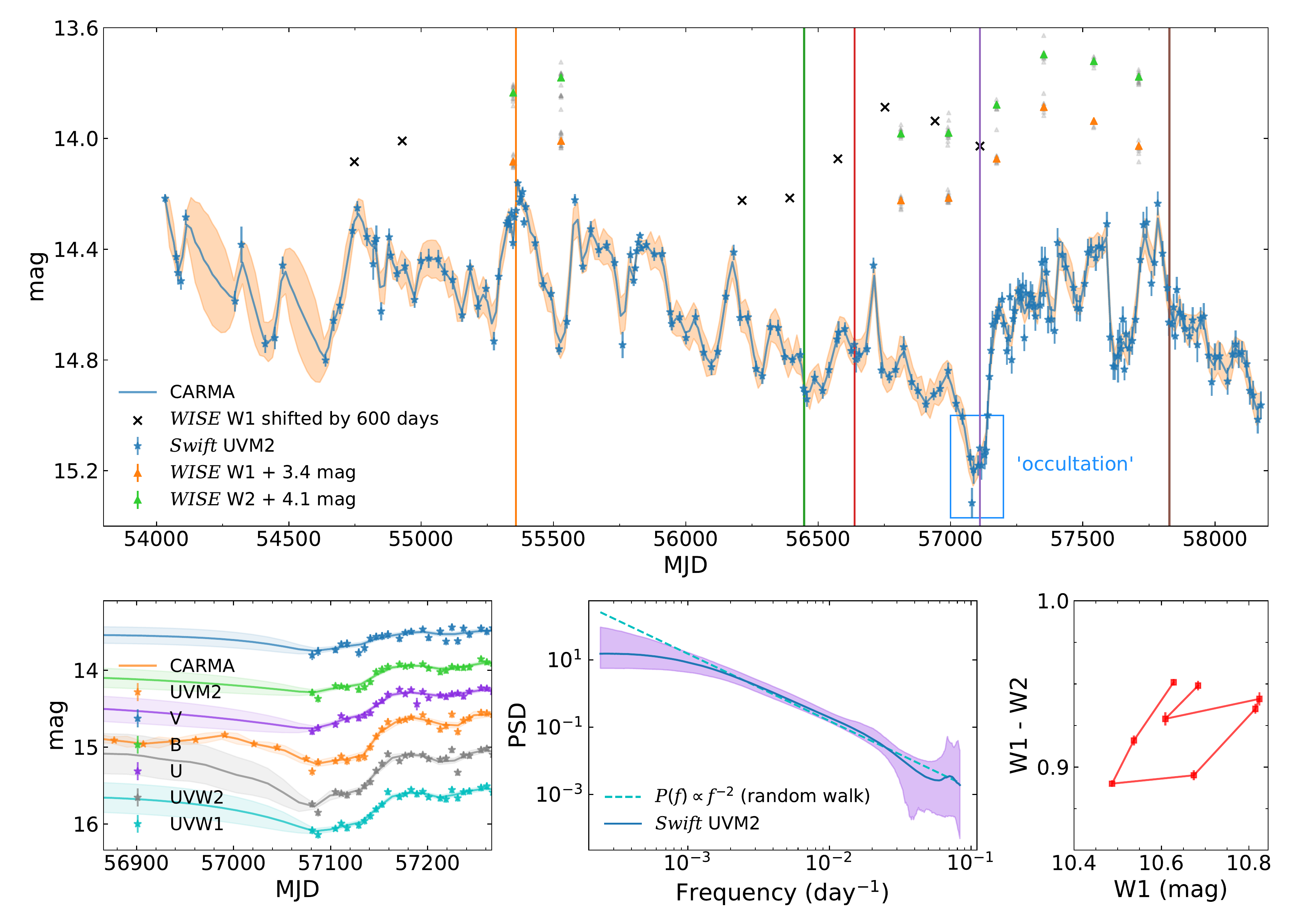}
\caption{{\it Upper}: \wise~and \swift~light curves for \wpvs. The time of five HST observations from 2010 to 2017 are labeled by solid lines. The \wise~data have been artificially shifted vertically for demonstration purpose. The WISE $W1$ data corrected for light-crossing time lag (by simply shifting the light curve by 600 days, see Section \ref{subsec:swift}) are shown in black crosses. The \uvm~light curve is fitted by the CARMA code \citep{Kelly2014} that models the AGN stochastic variability by a Gaussian continuous-time autoregressive process. The $1 \sigma$ confidence interval of the fit is shown in shaded region.  {\it Lower left}:  Zoom-in of the ``occultation event'' reported in \citealt{Leighly2015}. The data have been artificially shifted vertically for demonstration purpose. {\it Lower middle}: The solid line and shaded area represent the PSD and the corresponding 1 $\sigma$ confidence region of the \swift~\uvm~light curve derived from CARMA, respectively. The dashed line represents the PSD of the random walk model (i.e., $P(f) \propto f^{-2}$). {\it Lower right}: The $W1 - W2$ color as a function of $W1$ magnitude. The large \wise~variability amplitude and its color variability suggest that the intrinsic luminosity must change. The \swift~light curves including the occultation event (see Section \ref{subsec:cmp}) can be well fitted by the stochastic AGN variability model.
The PSD for \wpvs~has a high-frequency slope consistent with --2, and flattens at low frequencies, in agreement with the damped random walk model \citep[e.g.,][]{Kelly2009}.
These results suggest that the \swift~variability pattern is consistent with intrinsic AGN stochastic variability. }
\label{fig:lc}
\end{figure*}

The \wise\ mission \citep{Wright2010} and its successor, the \textit{Near-Earth\ WISE\ Reactivation} (\textit{NEOWISE}) mission \citep{Mainzer2014}, have imaged the full sky repeatedly since January 2010. The \wise~light curves for WPVS 007 at $W1$ and $W2$ bands are extracted from the IRSA service\footnote{See https://irsa.ipac.caltech.edu/Missions/wise.html.} with a 2$''$ matching radius. Data points with quality flags of $qi\_fact < 1$, $saa\_sep < 5$ and $moon\_masked = 1$ are excluded.\footnote{See https://wise2.ipac.caltech.edu/docs/release/allwise/expsup\\/sec3\_2.html for details.} The binned light curves are shown in Fig.~\ref{fig:lc} as orange ($W1$) and green ($W2$) triangles. The individual observations are shown in gray points.

\subsection{\textit{Swift} observations}
\label{subsec:swift}
WPVS 007 has been extensively monitored with \swift~\citep{Gehrels2004} using XRT and UVOT, including \uvww, \uvw, \uvm, \uu, \bb~and \vv~bands centered at 1928 \AA, 2246 \AA, 2600 \AA, 3465 \AA, 4392 \AA~and 5468 \AA, respectively \citep[e.g.,][]{Grupe2007, Grupe2008, Grupe2013}. We extract its \swift~light curves using the HEASOFT version 6.24 \textit{uvmaghist} tool. The source region is selected using a circle with a radius of $5''$ centered on the source and the corresponding background region is selected by an annulus with a $12.5''$ inner radius and a $25.0''$ outer radius. The binned \uvm~light curve (in bins of 6~days) is displayed in Figure \ref{fig:lc}.

The intrinsic 1450 \AA~luminosity ($L_{1450}$) is calculated by extrapolating the \uvw~band and \uu~band luminosities  with a power law. The photometry used in the calculation has been corrected for Galactic reddening assuming $E(B-V)=0.012$ mag \citep{Schlegel1998} and the \cite{Cardelli1989} extinction law. 
$L_{1450}$ is then converted into bolometric luminosity (\lbol) using ${\rm log}\,L_{\rm bol} = (4.745 \pm 1.007) + (0.910 \pm 0.022)\, {\rm log}\,\lambda L_{1450}$ \citep{Runnoe2012} and the resulting \lbol~for the six \textit{HST} observations (Section \ref{subsec:hst}) are presented in Table \ref{table:info}. 
The \uvw~and \uu~bands are utilized here because they are relatively less affected by possible intrinsic extinction (owing to their longer wavelengths compared to the \uvm~and \uvww~bands; e.g., \citealt{Calzetti2000}) and emission lines (note that the \bb~and \vv~bands are close to strong emission lines; see Figure \ref{fig:hst}). 
 Note that we mainly focus on the relative variability of luminosity in the following analysis. Therefore, we argue that the constant host galaxy contamination and a different bolometric conversion factor would not influence our main results materially.

\subsection{\hst~observations}
\label{subsec:hst}

\begin{figure*}
\centering
\includegraphics[width=\linewidth]{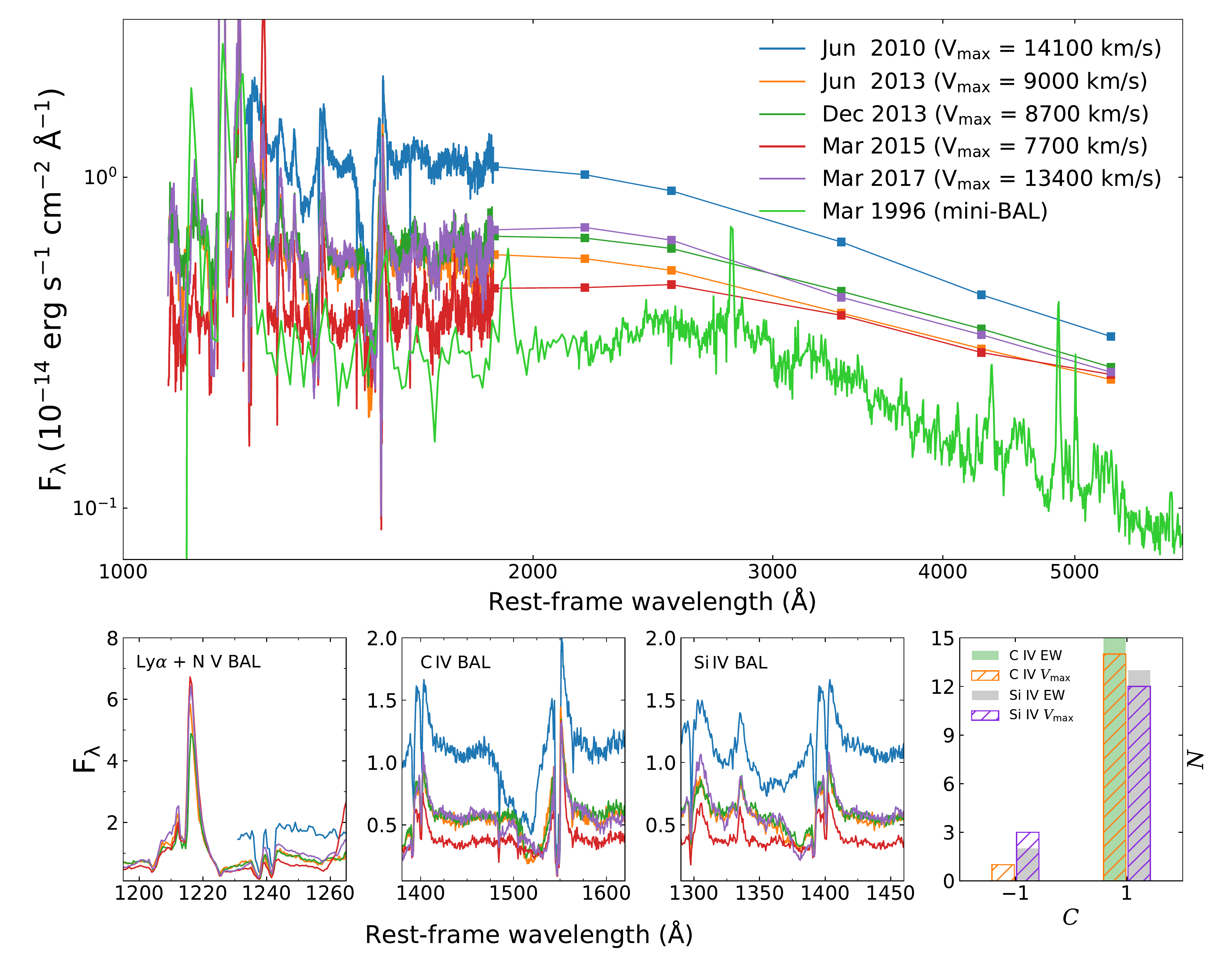}
\caption{{\it Upper}: \hst~spectra from 1996 to 2017. The \swift~photometries at the times of \hst~observations are shown as squares. {\it Lower left}: The regions of \lya, \nv, \civ,~and \siv~lines (The units are the same as the upper panel). {\it Lower right}: Distribution of concordance index $C$. $C$ is assigned as +1/$-1$ if the outflow strength (velocity) and bolometric luminosity vary in the same/opposite sign. The \civ~and \siv~BAL strength and velocity for most epochs vary coordinately with luminosity.}
\label{fig:hst}
\end{figure*}

\begin{table*}
\centering
\caption{Broad-absorption line properties of \wpvs~at the time of \hst~observations.}
\begin{tabular}{c c c c c c c} 
\hline
\hline
Date & \lbol~(\ergs) & \civ~$V_{\rm max}$ ($\rm km\ s^{-1}$) & \civ~EW (\AA) & \siv~$V_{\rm max}$ ($\rm km\ s^{-1}$) & \siv~EW (\AA)\\
\hline
1996 Jun & $5.2 \times 10^{43}$ & 900 (\nv) & -- & -- & -- \\
2010 Jun & $3.6 \times 10^{44}$ & 14100 & 20.4 & 11900 & 9.2\\
2013 Jun & $1.8 \times 10^{44}$ & 9000 & 15.1 & 6700 & 8.8\\
2013 Dec & $2.0 \times 10^{44}$ & 8700 & 15.9 & 6300 & 7.4\\
2015 Mar & $1.4 \times 10^{44}$ & 7700 & 6.3 & 4000 & 3.3\\
2017 Mar & $2.4 \times 10^{44}$ & 13400 & 19.2 & 5800 & 9.9\\
\hline
\end{tabular}

\vspace{10.0 pt}
{{\bf Note}. The bolometric luminosity and mini-BAL velocity for the 1996 \hst~observation are adopted from \cite{Leighly2009} and \cite{Leighly2015}, respectively}.
\label{table:info}
\end{table*}

\begin{figure*}
\centering
\includegraphics[width=0.49\linewidth]{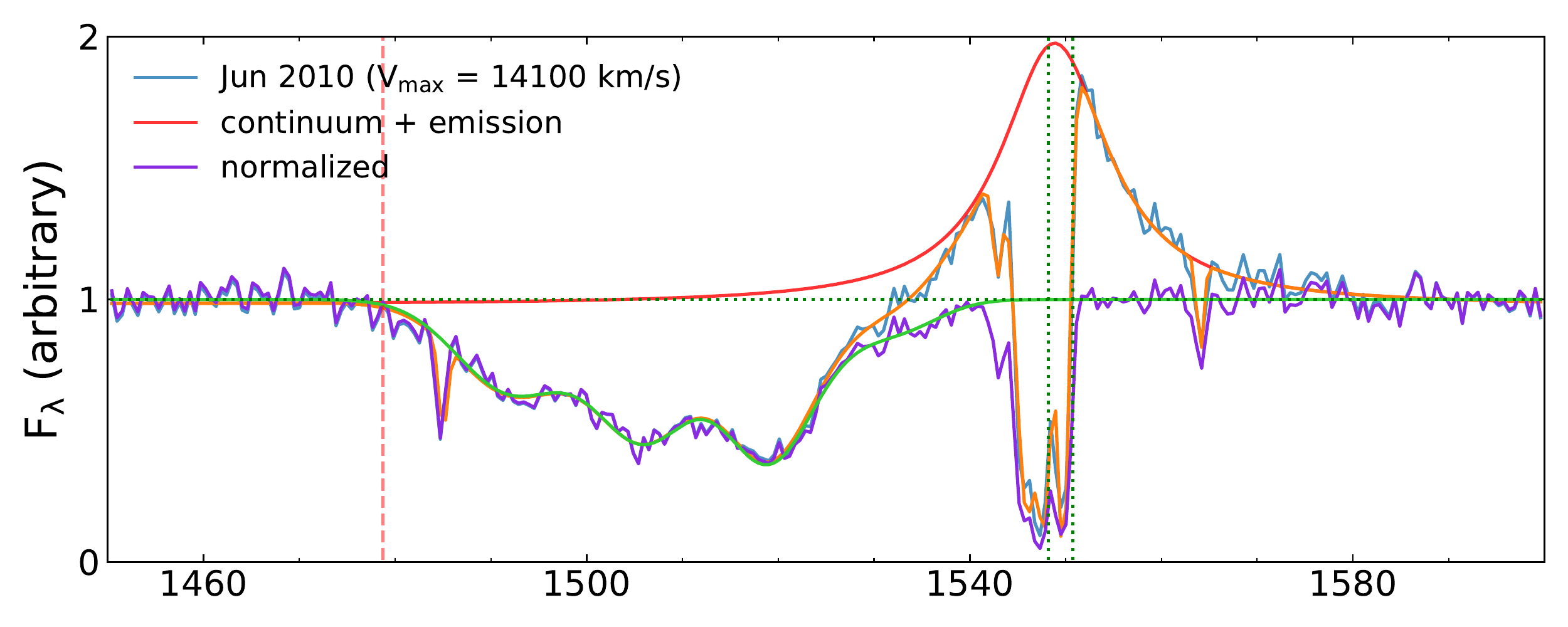}
\includegraphics[width=0.49\linewidth]{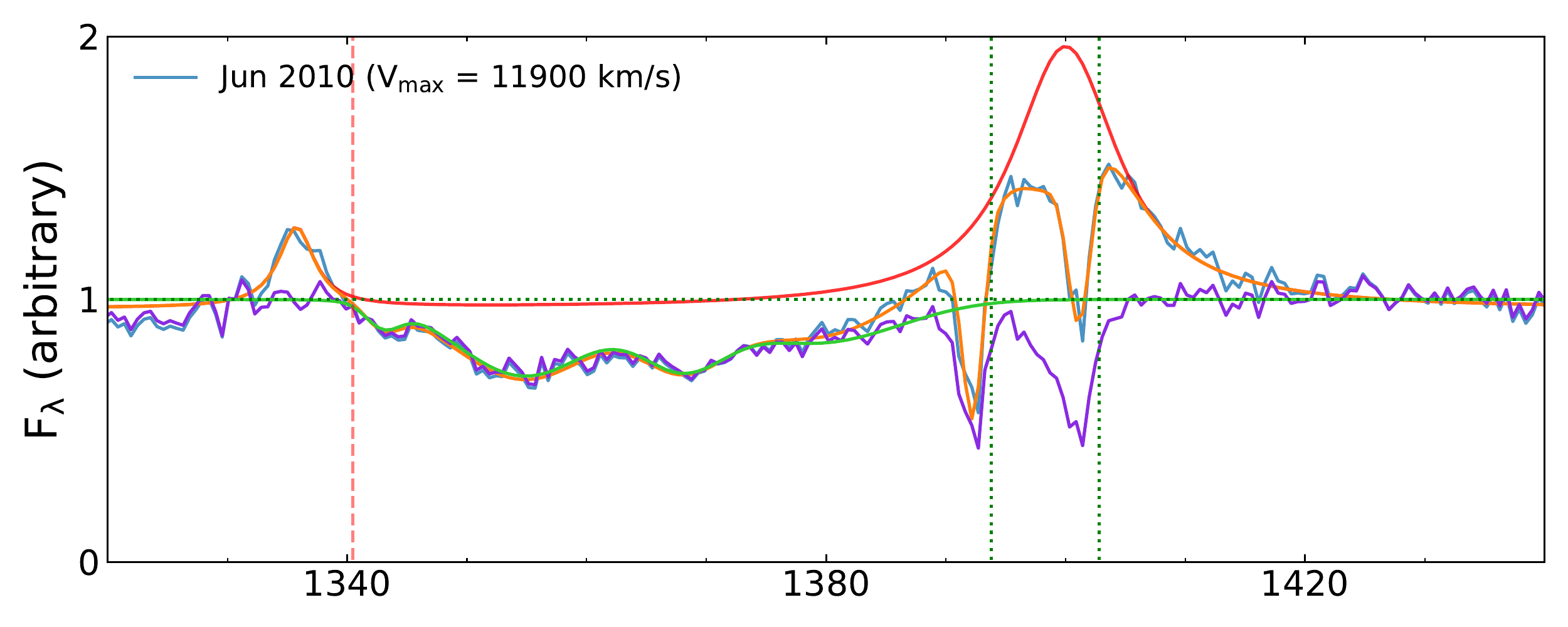}

\includegraphics[width=0.49\linewidth]{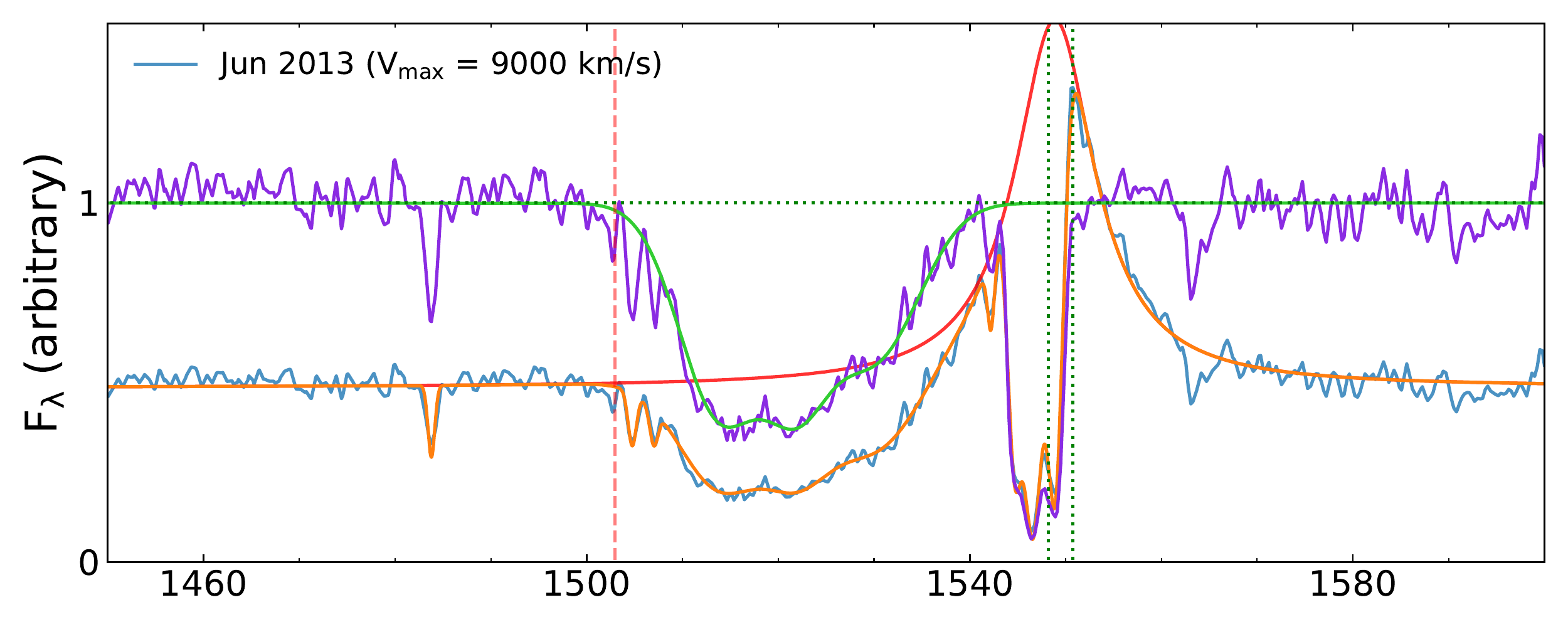}
\includegraphics[width=0.49\linewidth]{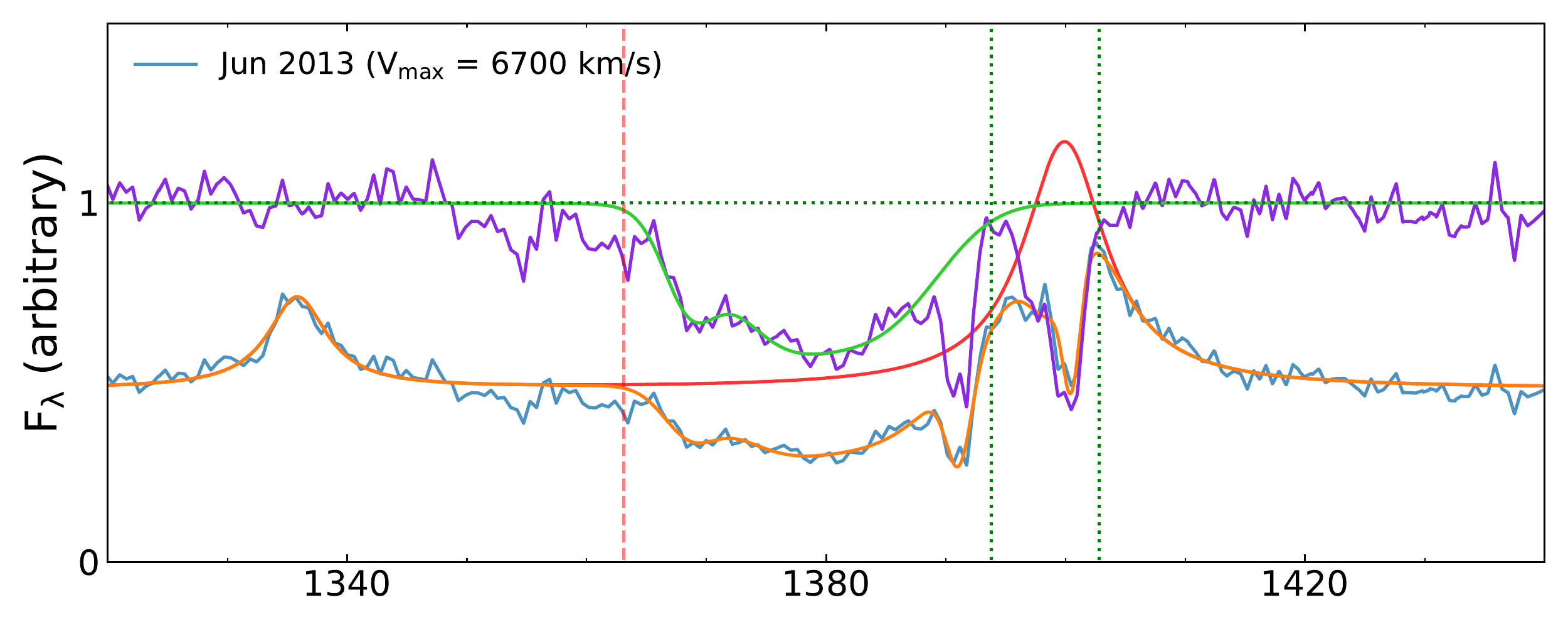}

\includegraphics[width=0.49\linewidth]{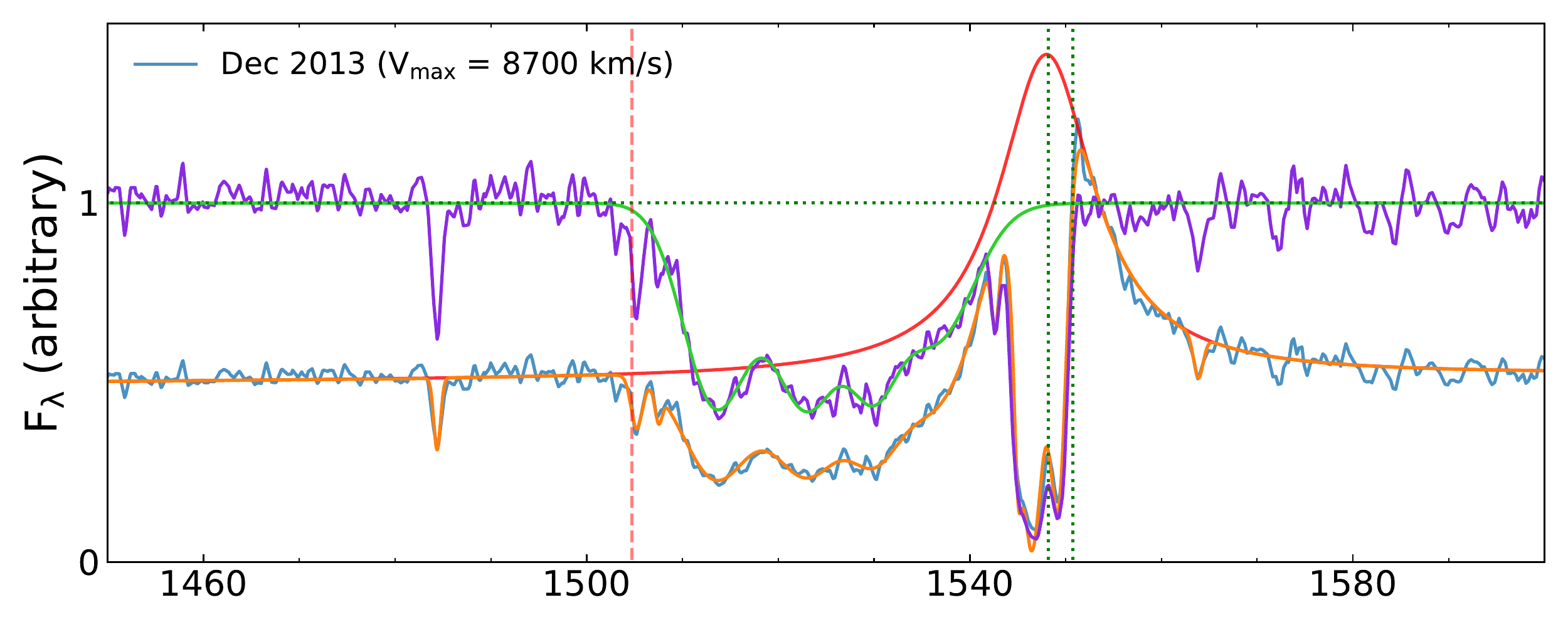}
\includegraphics[width=0.49\linewidth]{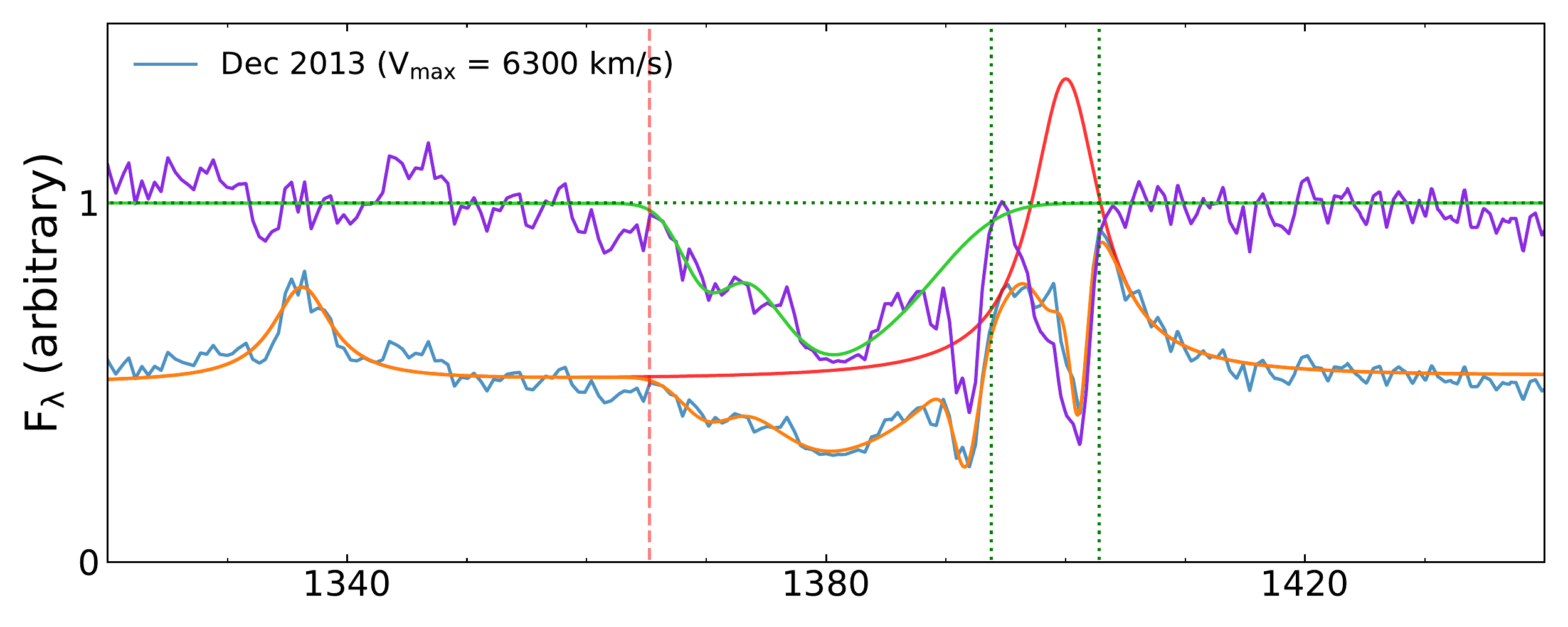}

\includegraphics[width=0.49\linewidth]{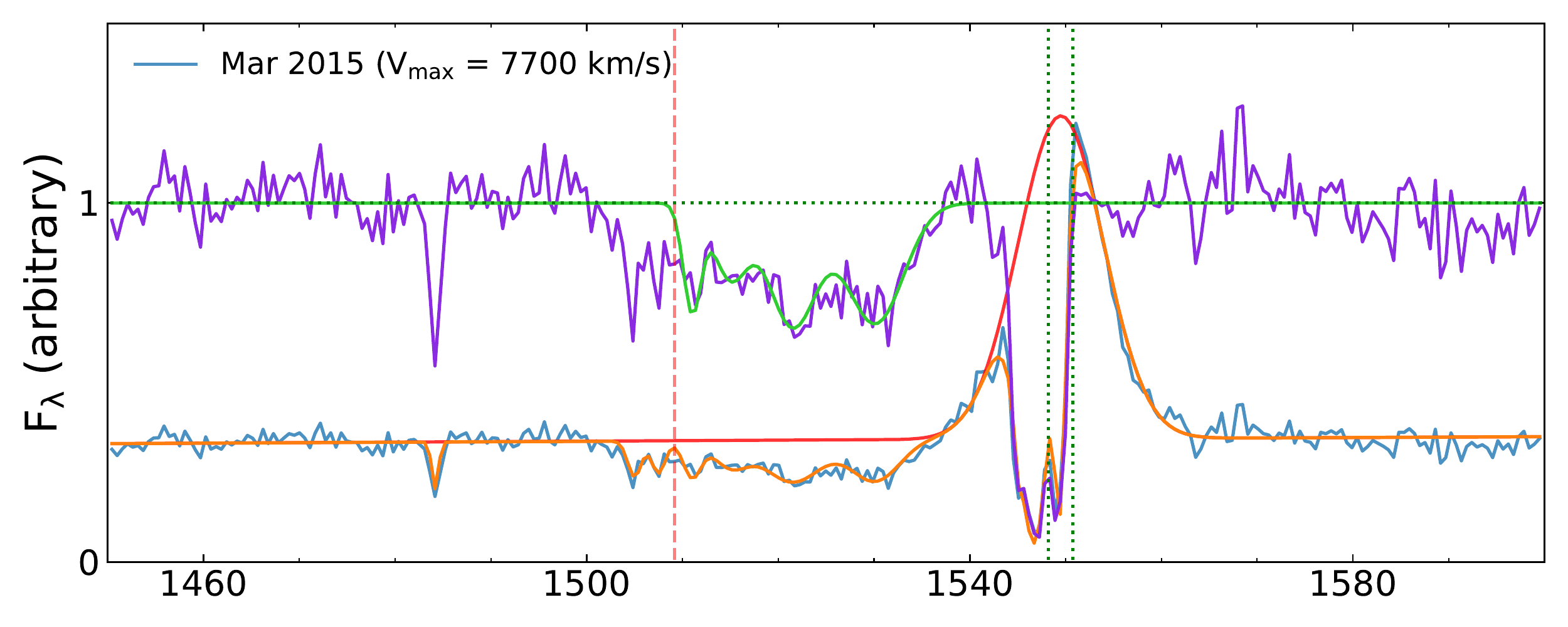}
\includegraphics[width=0.49\linewidth]{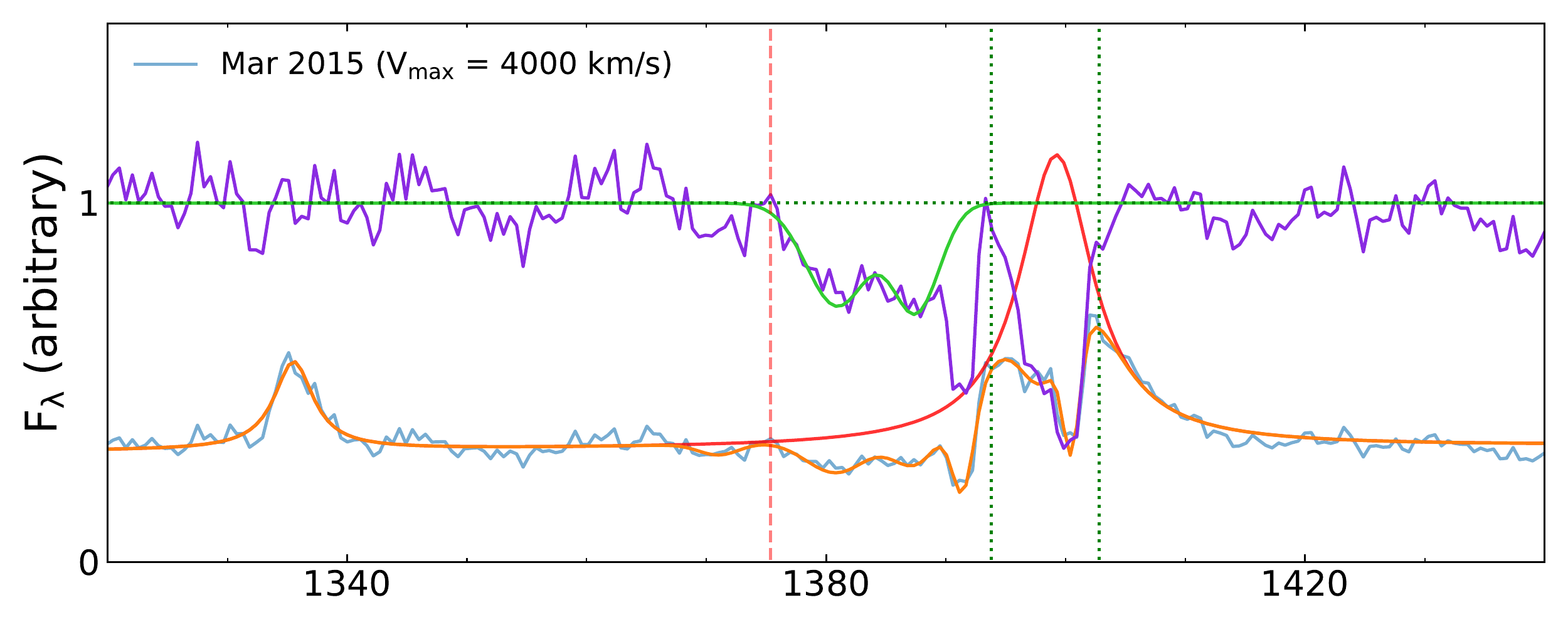}

\includegraphics[width=0.49\linewidth]{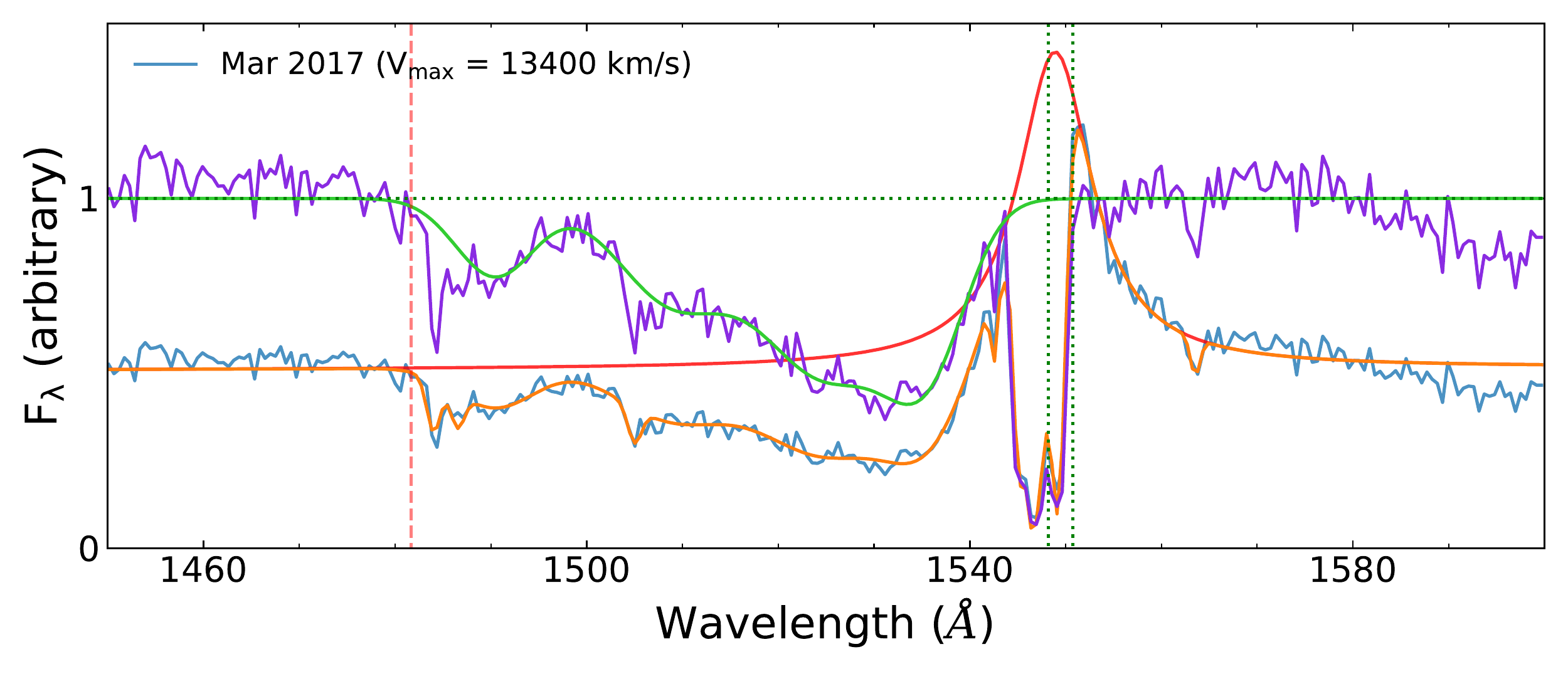}
\includegraphics[width=0.49\linewidth]{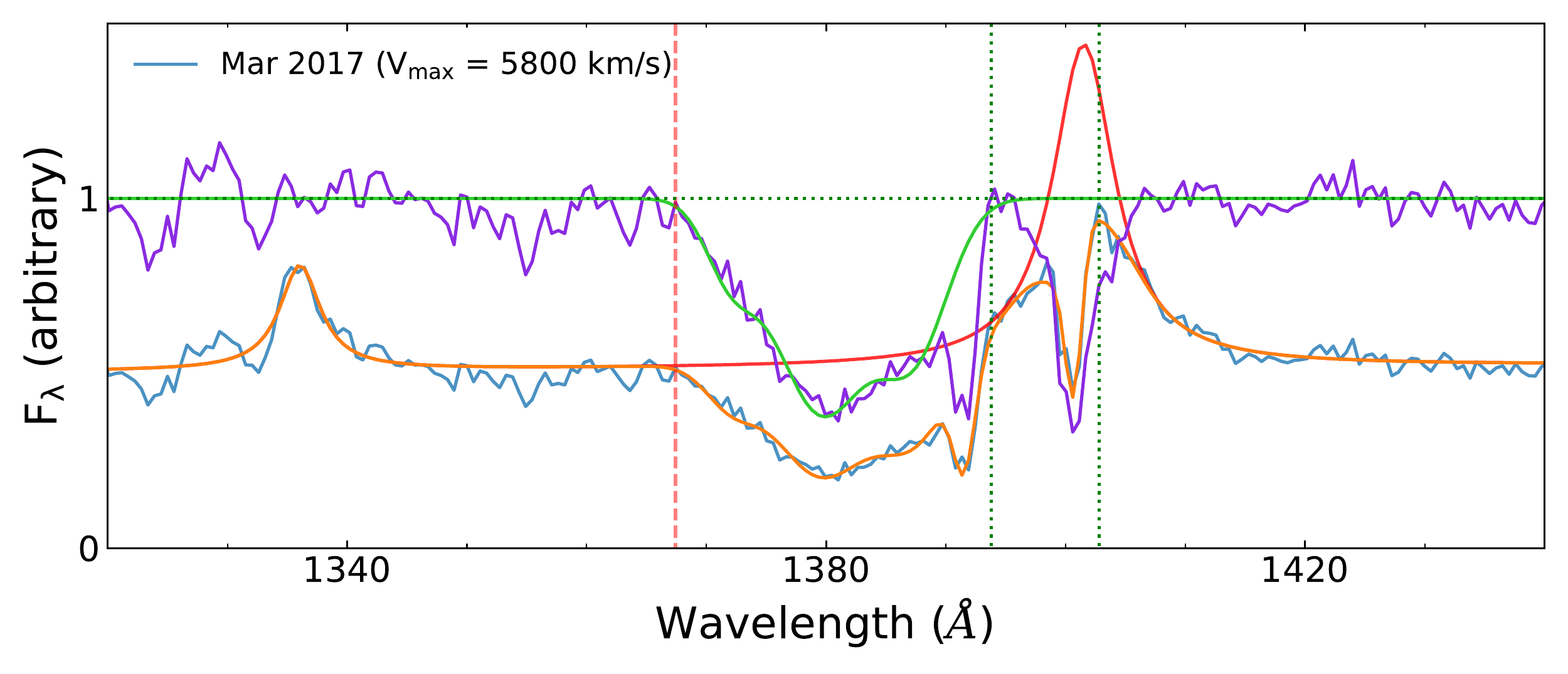}

\caption{Spectral fitting results of \civ~(left) and \siv~(right) BAL regions. 
The observed spectrum is shown in blue. The spectral fitting result is shown in orange. The best fit emission line plus continuum is shown in red. The purple curve represents the normalized spectrum. The normalized model-fitting BAL component is shown in green.  
The dotted lines at (1548.2 \AA, 1550.8 \AA)~and (1398.8 \AA, 1402.8 \AA)~represent the rest-frame wavelengths of \civ~and \siv~doublets, respectively. The red vertical dashed line marks the position of the observed maximum outflow velocity defined in Section \ref{subsec:hst}.}
\label{fig:civ}
\end{figure*}

WPVS 007 has been observed six times by the Faint Object Spectrograph (FOS) or Cosmic Origins Spectrograph (COS) aboard \hst~since 1996 (see Table \ref{table:log}). The spectra are extracted from the \hst~online dataset\footnote{See http://archive.stsci.edu/hstonline.} that are calibrated and shown in Figure \ref{fig:hst} after smoothing by averaging the adjacent seven pixels. The BAL regions of \civ, \siv~and \nv~are plotted in bottom panels. 
To obtain the BAL velocity, we fit the observed spectra by an empirical model. We model the continuum with a power law in line-free regions (1000--1430 \AA, 1470--1610 \AA, 1620--1680 \AA, 1740--1760 \AA~and 1850--2900 \AA). 
The absorption features are fitted using a set of Gaussian profiles. The emission lines are modeled with Lorentz profiles except for the 2015 spectrum for which we use the Gaussian profile since the {\tt{curve\_fit}} python package we used is unable to find a solution with the Lorentz profile. The BAL troughs are determined as a combination of broad Gaussians, and the observed maximum velocity of the outflow is calculated as the velocity corresponding to 
the first point of the best fitted normalized BAL profiles which equals to 0.99 counted from the bluest wavelength.
The derived velocities of \wpvs~are summarized in Table \ref{table:info} and the spectral fitting results of \civ~and \siv~BAL regions are shown in Figure \ref{fig:civ}.

\section{Change of obscuration or intrinsic luminosity?}
\label{sec:variability}

\subsection{\wise~variability} 
\label{subsec:wise}
MIR variability has been widely utilized as a powerful tool to distinguish various variability mechanisms. The AGN MIR emission in the \wise~$W1$ and $W2$ bands is mainly produced by the hot dust re-radiation surrounding the central engine. Due to the small extinction in the MIR bands, large-amplitude MIR variability should only arise from intrinsic luminosity variability rather than the change of obscuration \citep{Sheng2017, Stern2018}. For \wpvs~that exhibits $\sim 0.3$ mag variability in the \wise~bands (see Figure \ref{fig:lc}), it would require a $\Delta A_{\rm V} \sim 5$ change in extinction if a variable obscuration scenario is assumed, insensitive to the adopted extinction models \citep{Cardelli1989, ODonnell1994, Fitzpatrick1999}. This would transform into a $\sim 5$ mag change in the \vv~band and $\sim 14 - 16$ mag change in the \uvm~band, which are all significantly larger than the true variability amplitudes during the \swift~monitoring (i.e., only $\sim 1$ mag change at maximum). Moreover, we find that the $W1 - W2$ color generally increases with increasing $W1$ magnitude (see the inset in Figure \ref{fig:lc}), suggesting that the MIR color becomes redder when the source dims. This might be explained by that the torus temperature decreases with decreasing luminosity, thus the spectral energy distribution (SED) becomes redder in the faint state. Such a phenomenon is also unexpected if the UV-optical flux variabilities are dominated by the obscuring materials move into/out-of the sightline. Therefore, the MIR variability observed in \wpvs~suggests that its intrinsic luminosity must vary significantly.

\subsection{Color variability}
\label{subsec:color}
As the intrinsic luminosity changes, the color would also change in response and is typically observed as being ``bluer-when-brighter'' that has been widely found in quasar variabilities \citep[e.g.,][]{Ruan2014, Guo2016}, which could be well explained by the stochastic temperature fluctuation in an inhomogeneous accretion disk \citep[e.g.,][]{Dexter2011, Cai2016, Cai2018}. Alternatively, the flux variability caused by variable extinction also displays a similar trend, since extinctions are more significant at shorter wavelengths \citep[e.g.,][]{Calzetti2000}. 

\begin{figure*}
\centering
\includegraphics[width=0.32\linewidth]{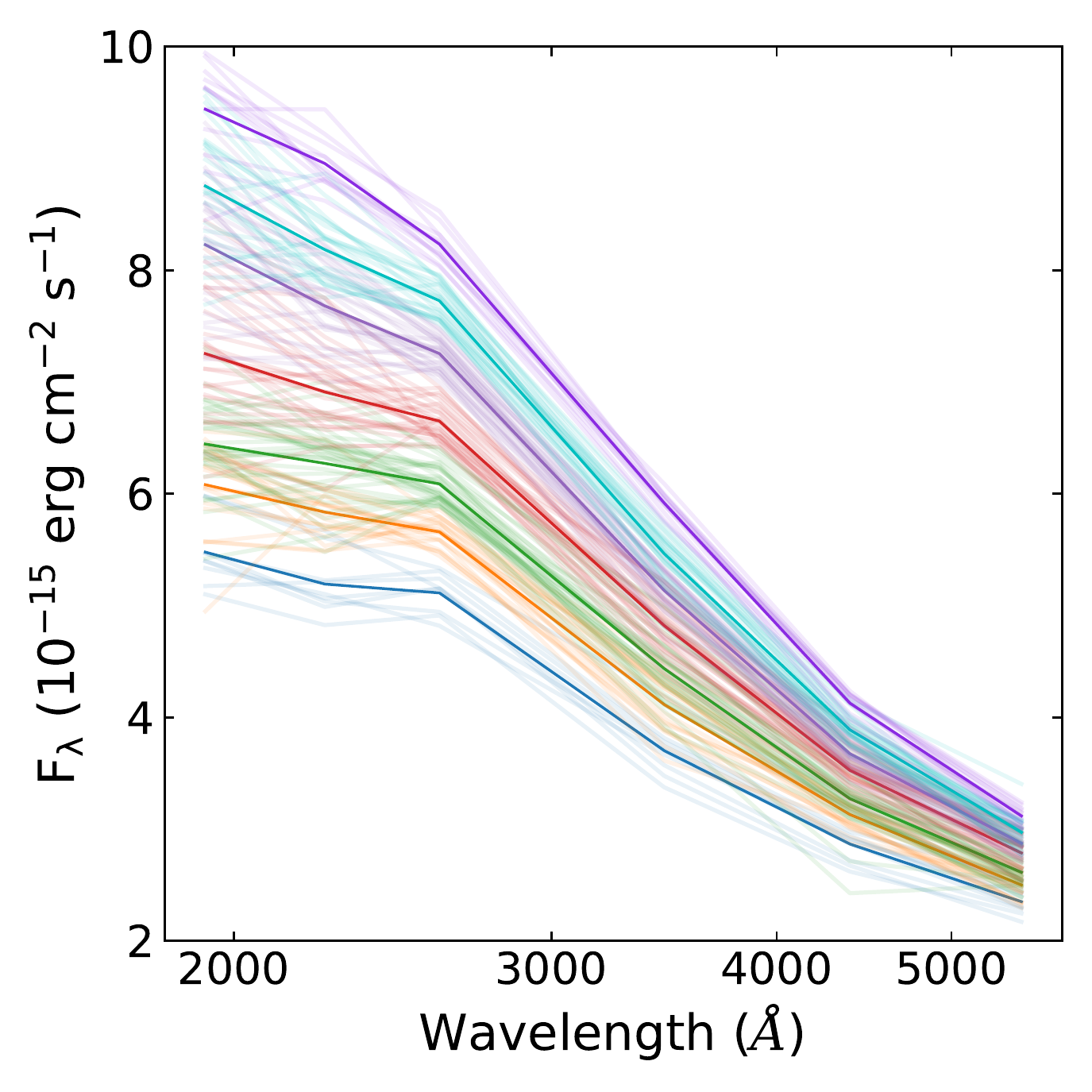}
\includegraphics[width=0.32\linewidth]{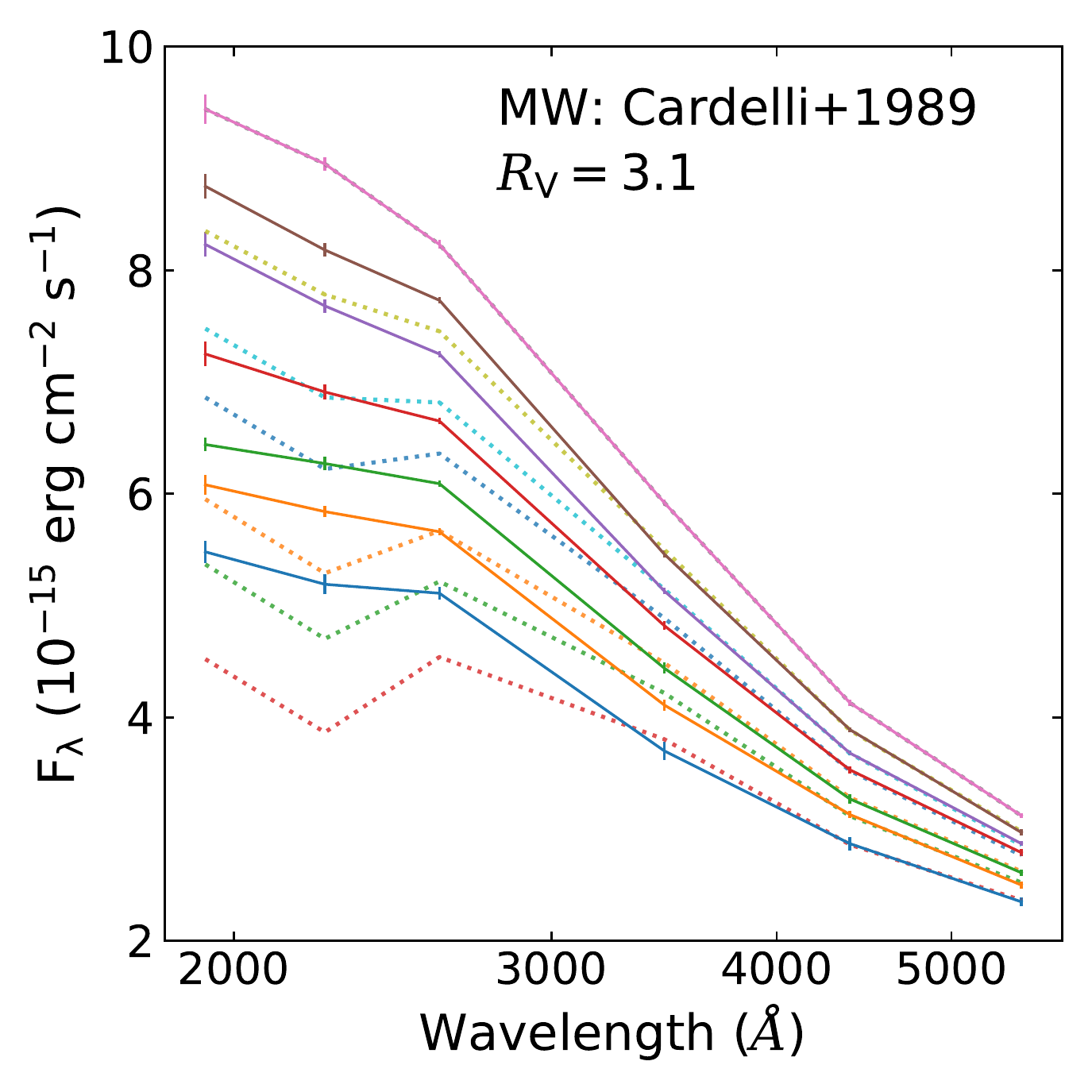}
\includegraphics[width=0.32\linewidth]{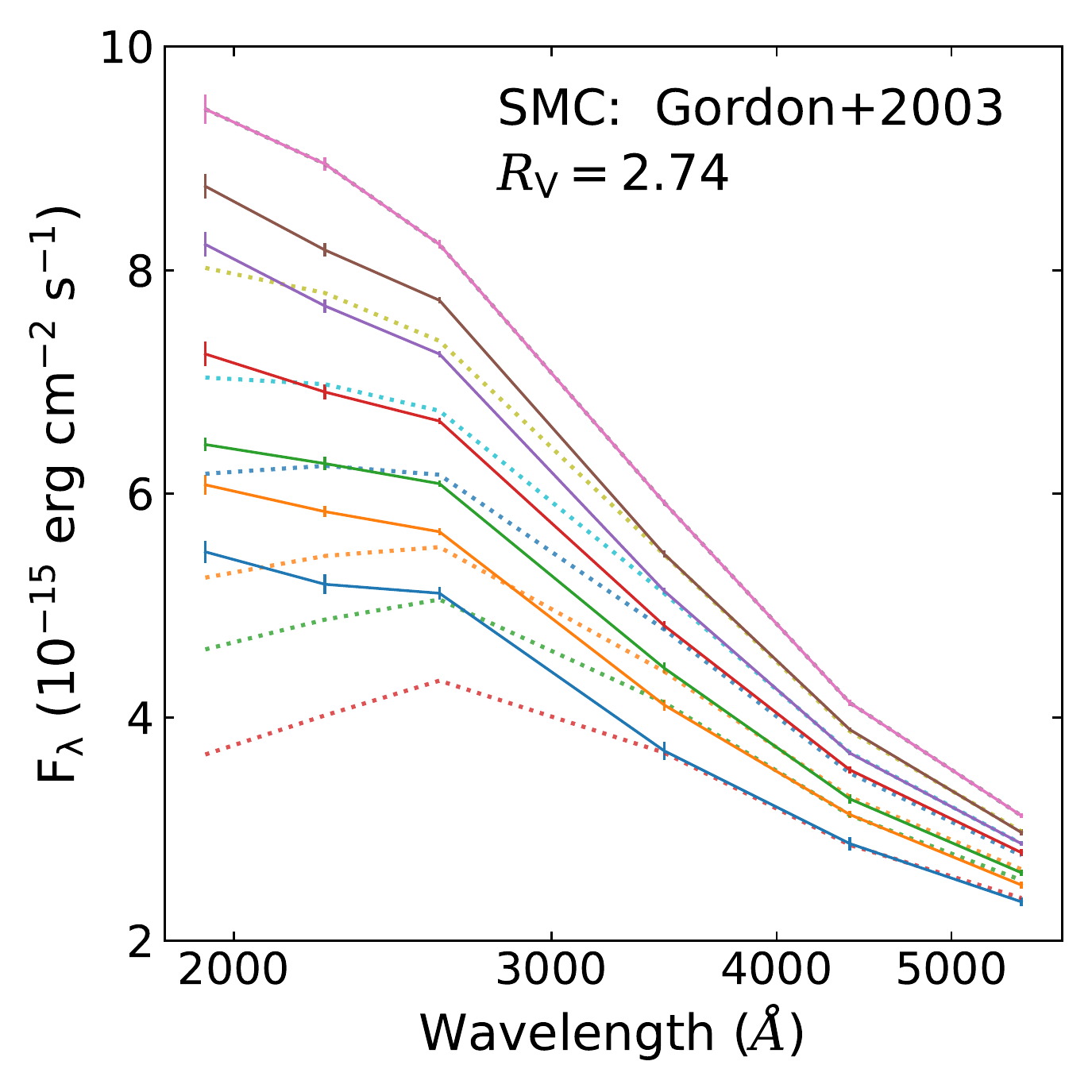}

\caption{{\it Left}: \swift~SEDs color-coded and sorted by their \uvw~fluxes. The individual SEDs are shown in light colors while the average SEDs in seven \uvw~flux bins are shown in bold curves. A clear bluer-when-brighter trend can be seen.  {\it Middle}: The solid curves are the average \swift~SEDs that are the same as those in the left panel. 
The dotted curves represent the SEDs derived from extincting the top SED (i.e., with the highest flux) 
using the \citealt{Cardelli1989} MW extinction law with $A_V$ ranging from 0 to 0.3. 
{\it Right}: Same as the middle panel, but using the \citealt{Gordon2003} SMC extinction law. 
The pure variable extinction model fails to reproduce the observed color variability, since it significantly underestimates UV fluxes for a given $A_V$ although the extincted SEDs largely match the observed SEDs at longer wavelengths.}
\label{fig:color}
\end{figure*}

To examine whether the color variability in \wpvs~also requires the change of intrinsic luminosity, we divide \swift~light curves into seven bins based on their \uvww~fluxes. The SEDs in each bin and their average values are shown in Figure \ref{fig:color}, in which the bluer-when-brighter behavior can be clearly seen. If variable extinction governs the color variability, we should be able to reproduce low-flux SEDs by simply adding dust extinction to high-flux SEDs. Note that L15 captured a particular occultation event lasting for $\sim 60$ days in 2015 and explained it as a result of variable extinction (but see Section 4.4). Since our \wise~data cannot deny this possibility, we exclude the data during the occultation event while calculating the SEDs (but we note that adding them back does not affect our results).

We test various extinction models \citep{Cardelli1989, ODonnell1994, Fitzpatrick1999, Gordon2003} and show the results of using the Milky Way \citep[MW; ][]{Cardelli1989} and Small Magellanic Cloud \citep[SMC; ][]{Gordon2003} extinction curves in Figure \ref{fig:color} for illustration. The top SED is extincted using a range of $A_V$ values from 0 to 0.3. For the resulting SEDs at given $A_V$ values (shown as dotted curves) which match well with the observed SEDs at $B$ and $U$ bands, they all significantly underestimate the fluxes at \uvm~and \uvw~bands, insensitive on the assumed values of $R_V=A_V/{\rm E}(B-V)$. 
Note that we do not consider the \uvww~band here since it might be contaminated by the \ciii~emission line. 
This result suggests that the variable extinction model fails to reproduce the bluer-when-brighter trend, and intrinsic variability is required to account for the observed color variability of \wpvs.

\subsection{\swift~variability}
\label{subsec:swift}

As discussed in Sections \ref{subsec:wise} and \ref{subsec:color}, the intrinsic luminosity of \wpvs~must vary significantly to account for the long-term \wise~and UV to optical color variations, and the change of obscuration may play a subordinate role. If our model is correct, we should be able to explain the flux variations in UV-optical bands by means of stochastic changes of intrinsic AGN power, and the corresponding  power spectral density (PSD) should be similar to those of other type 1 AGNs. To examine the variability pattern of \swift~light curves, we fit them with the CARMA code\footnote{See https://github.com/brandonckelly/carma\_pack.} \citep{Kelly2014} and derive the power spectral densities (PSDs). The CARMA code models the AGN stochastic variability using a Gaussian continuous-time autoregressive process,
and uses two order parameters, $p$ and $q$, to describe autoregressive coefficients and moving average coefficients of different stochastic processes.
We determine $p$ and $q$ using the {\tt{choose\_order}} method in CARMA that selects the best model using the Akaike Information Criterion (see Section 3.5 in \citealt{Kelly2014} for details). The fitting results and the PSD are shown in Figure \ref{fig:lc}.  The PSD for the random walk model (i.e., $P(f) \propto f^{-2}$) is also displayed for comparison.

It is evident that the ten-year \swift~light curves can be well fitted by a $p = 7$, $q = 2$ CARMA process (note that we only display the fitting result for the \uvm~band but light curves in other bands can be also well fitted), including the occultation event reported in L15 (see our Section \ref{subsec:cmp}). 
Comparing the derived PSD shape with that of the random walk model, we find that the high-frequency PSD slope for \wpvs~is   consistent with --2, i.e, the slope of the random walk process. The PSD then flattens at low frequencies, suggesting that the variabilities are damped at longer timescales \citep[e.g.,][]{Kelly2009}. 
Such a PSD shape is similar to those of other type 1 AGNs reported in previous studies \citep[][]{Kelly2009, MacLeod2010, Zu2013, Sun2015}, and might be caused by
the thermal fluctuations in the accretion disk \citep[e.g.,][]{Kelly2009, Sun2018}.
Therefore, our results suggest that the intrinsic stochastic AGN variability being caused by a fluctuated accretion disk is able to well reproduce the observed \swift~variability behavior, without the necessity to invoke other significant variable factors such as the change of dust extinction. 

In addition, based on the 5100 \AA~luminosity ($\sim 3.5 \times 10^{43}\ \ergs$ on Jun 2010), we estimate the inner-torus sublimation radius of \wpvs~to be $\sim 0.08$ pc \citep{Netzer2015}; the torus size at $\sim 8-13\ \um$ is estimated to be $\rm \sim1\ pc$ based on the relationship obtained through MIR-interferometry observations (see Figure 4 in \citealt{Netzer2015}). Therefore, the torus size at \wise~bands ($\sim 5\ \um$) might lie between $\sim 0.1 - 1\ \rm pc$. Converting the distance into light crossing time, we expect the time delay between \wise~and \swift~variabilities to be $\sim 600\rm\ days$ by rougly adopting $R_{4.6\ \um} = 0.5\ \rm pc$.  Although the sparse sampling of  \wise~observations does not allow us to robustly constrain the time delay, simply shifting \wise~light curves by 600 days yields good agreement between \wise~and \swift~profiles, i.e., the rise and later decline of \wise~data are consistent with the 
brightening and dimming of \swift~data between MJD 56300 to MJD 56700, and MJD 56700 to MJD 57000, respectively (see the black crosses in Figure \ref{fig:lc}), thereby lending further support to the intrinsic nature of UV-optical variabilities.

\section{Discussion}
\label{sec:discussion}
Our analysis of archival \wise~and \swift~data suggests that the large-amplitude MIR variability, color variability, and UV-optical continuum flux variabilities discovered in the ten-year monitoring of WPVS 007 are most probably driven by the stochastic change of intrinsic luminosity,  and variable extinction may play a minor role.

\begin{table}
\centering
\small
\caption{Thermal, viscous, and heating-cooling front timescales at \textit{Swift} bands.}
\begin{tabular}{c c c c} 
\hline
\hline
$\lambda$ (\AA) & $t_{\rm thermal}$ (yrs) & $t_{\rm viscous}$ (yrs) & $t_{\rm front}$ (yrs)\\
\hline
1929 & 0.011 & 1.68 & 0.14\\
2242 & 0.015 & 3.34 & 0.23\\
2609 & 0.021 & 6.48 & 0.37\\
3460 & 0.038 & 23.78 & 0.95\\
4384 & 0.061 & 68.75 & 2.05\\
5472 & 0.096 & 1769.30 & 4.14\\
\hline
\end{tabular}
\label{table:timescale}
\end{table}

Following Section 3.2 (Page 108) of \cite{Kato2008}, we calculate the timescales that are relevant to various intrinsic variability mechanisms at different wavelengths (radii): the thermal timescale $t_{\rm th}$, which represents the timescale of thermal fluctuation of accretion disk; the viscous timescale $t_{\rm vis}$, on which the global accretion rate changes; and the heating-cooling front timescale $t_{\rm front}$, corresponding to the heating and cooling fronts propagating through an $\alpha$ disk (i.e., the viscosity in the accretion disk, which is crucial for driving the angular momentum of the accreted materials outward, is assumed to be $\alpha p$, where $\alpha$ and $p$ are the dimensionless viscous parameter and the total pressure, respectively; see \citealt{SS1973}). The results are listed in Table \ref{table:timescale}. Note that these values are derived under the assumption of a standard thin disk model for simplicity, since for $\edd \sim 0.4$ as in WPVS 007, the disk remains geometrically thin at NUV/optical wavelengths. The local thermal fluctuation with $t_{\rm th}\sim$ a few days produces the seed of short-timescale variability at a given radius \citep[e.g.,][]{Sun2018}. The $t_{\rm front}$ falls within the observed timescale, which makes it possible that the inner accretion disk could go through significant overall variability as the cooling and heating fronts propagate radially across an $\alpha$ disk, thus causing large variations at multiple wavelengths \citep[e.g.,][]{Stern2018, Ross2018}.

In the intrinsic luminosity variability scenario and within the radiatively-driven outflow framework, the outflow properties may display dramatic variability as the incident ionizing continuum changes. In the following sections, we investigate the BAL variability in WPVS 007, aiming at testing whether its behavior is consistent with being photoionization-driven, and try to put constraint on the outflow launch radius \citep[e.g.,][]{Parker2017, Pinto2018}.

\subsection{Change of ionization state and BAL trough strength}
\label{subsec:ionization}
The ionization parameter ($U \propto \frac{L}{n_{\rm H} R^2}$) and ion column densities of the outflowing gas would change in response to the variable ionizing luminosity, resulting in the change of BAL strength.
Using the best fit normalized BAL component (see the green curves in Figure \ref{fig:civ}), we calculate the equivalent width (EW) of \civ~and \siv~BALs and list the results in Table \ref{table:info}. To qualitatively describe its relationship with luminosity, we calculate the concordance index $C$ following \cite{Wang2015}. 
$C$ is assigned as +1 if BAL strength and bolometric luminosity vary in the same sign (i.e., both increasing or decreasing); otherwise, it is assigned as $-1$. The distribution of $C$ for any two observations is shown in the bottom-right panel of Figure \ref{fig:hst}. All (most) epochs have $C = 1$ for \civ~(\siv) BAL, which suggests that the BAL and luminosity of \wpvs~weaken and strengthen simultaneously.

Such a trend is different from what is typically observed in luminous quasars, for which the weakening of BAL trough is often accompanied with the brightening of continuum \citep[e.g.,][]{Wang2015, He2017, Pinto2018}. This may be explained by the different ionization states between \wpvs~and quasars. 
For \wpvs, \cite{Leighly2009} derived $\nh \sim \cm$ (see also \citealt{Grupe2013}) and ${\rm log}\,U \gtrsim -0.3$  from the \fuse~spectrum. The persistently weak \nv~line during 2010 to 2017 (see Figure \ref{fig:hst}) also suggests that it is in a lower ionization state compared with other luminous BAL quasars, possibly related to its low luminosity and X-ray weak nature \citep{Grupe2007, Grupe2008, Grupe2013, Gibson2008}.
As shown in Figure 8 in \cite{He2017}, for $\nh \sim \cm$, the column densities of \civ~and \siv~ions reach the peaks at ${\rm log}\, U \sim 0$ and monotonically decrease at both sides. 
Therefore, for the low-ionization source \wpvs~that lies on the left side of the peak of the $N_{\rm ion}$ vs. ${\rm log}\,U$ curve, the BAL strength and luminosity vary in a coordinated way. 

\subsection{Coordination between outflow velocity and AGN luminosity}\label{subsec:velocity_lumin}
We use the same concordance index method introduced in Section \ref{subsec:ionization} to describe the relationship between BAL blue-shifted velocity and source luminosity. A  similar coordinated correlation between bolometric luminosity and the ``observed'' maximum velocity can be also seen from Table \ref{table:info} that \vmax~increases with increasing \lbol, including the faintest state in 1996 when no BALs were detected. For the \civ~BALs, $C$ is 1 for 14 out of 15 epochs, and the only exception happened at the two 2013 observations, at which the luminosity and velocity for the two epochs are actually very close to each other. While for the \siv~BALs, $C$ is 1 for 80\% (12/15) of the observations.

We note that if the change of apparent velocity is caused by a flowing structure randomly crossing our sightline \cite[e.g.,][]{Hamann2008}, equal probabilities of concordance (i.e., $C=1$) and anti-concordance (i.e., $C=-1$) between \vmax~and \lbol~should be expected in each pair of observations. Therefore, considering all observations, we should expect that half of the observations have $C=1$ while the remaining have $C=-1$ (i.e., the distribution of $C$ should follow a binomial distribution with a concordance probability of being 0.5). However, we find that 14 (12) out of 15 observations for \civ~(\siv) BALs have $C=1$. Under the assumption of the moving-cloud scenario, the binomial probabilities for observing such events are only 0.0005 for \civ, and 0.014 for \siv, respectively. Such low probabilities suggest that it is not very likely that the change of observed maximum velocity is caused by the transverse motion of  obscuring clouds.

Instead, the coordinated correlation can be easily explained if the BAL variability is photoionization-driven \citep[e.g.,][]{Hamann2011, FilizAk2012, FilizAk2013, Wang2015, He2017}. As pointed out in Section \ref{subsec:ionization}, the \civ~and \siv~column densities decrease when the source becomes fainter. For the high-velocity components of the outflow that typically have small absorption strengths and are more likely to vary \citep[e.g.,][]{Capellupo2012}, they would become weaker and disappear in the spectrum when the optical depth reduces, leading to the decrease of the ``observed'' maximum velocity when the source dims. 
The component with $V_{\rm max} > 10000\rm \ km\ s^{-1}$ emerged for the \civ~BAL in the Mar 2017 observation but no corresponding component was found for the \siv~BAL (see Figure \ref{fig:civ}) suggests that, the smaller optical depth of \siv~is responsible for the absence of the high-velocity component in the \siv~trough, owing to much smaller element abundance of \siv~compared to \civ.
In addition, the increasing radiation pressure from Mar 2015 to Mar 2017 may also contribute to the increasing velocity, but current data do not allow measurement of the actual acceleration.

\subsection{BAL velocity and outflow launch radius}
\label{subsec:torus}

\begin{figure*}
\centering
\includegraphics[width=0.7\linewidth]{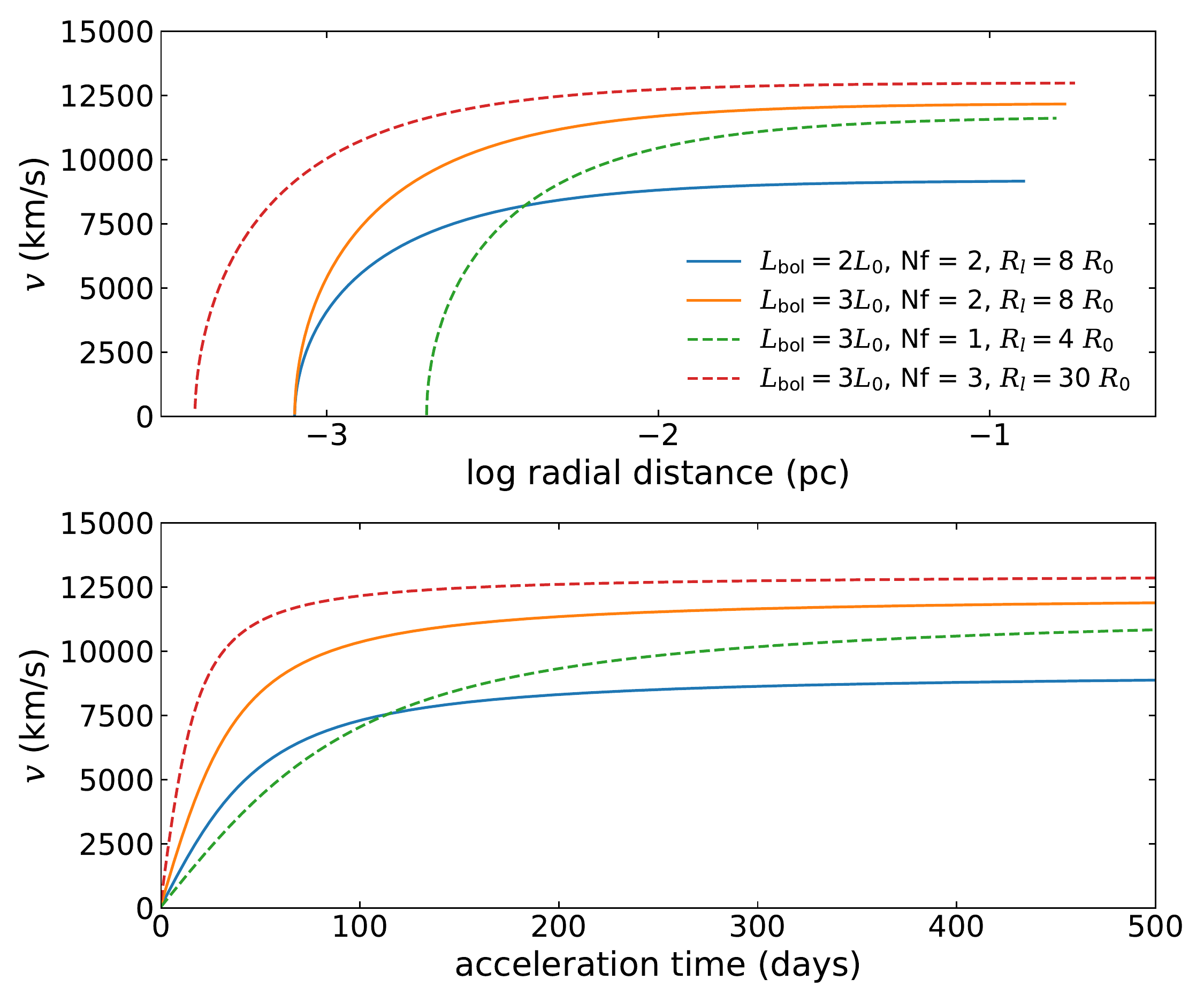}
\caption{{\it Top}: Outflow radial velocity as a function of radial distance for different initial conditions calculated from Eq.~\ref{eq:motion} (where $L_0$ = $\rm 10^{44}\ \ergs$, Nf = $N_{22} / f_{0.1}$, and $R_0 = \rm 10^{-4}\ pc$). The initial conditions for curves in the same colors in both panels are the same. A clear correlation between the maximum outflow velocity and luminosity can be seen. For outflows launched in the inner region, e.g., $R_l = \rm 0.0008\ pc$, the observed outflows that have $v \sim 10000\rm\ km\ s^{-1}$ may be located far away from its birth region and lie close to the torus ($\sim 0.1$ pc), since they propagate and are accelerated outward. {\it Bottom}: Outflow radial velocity as a function of acceleration time. For the launch radius as small as 0.0008 pc, the outflow soon reaches its maximum velocity in the observed timescales.} 
\label{fig:velocity}
\end{figure*}

If the outflows are launched radiatively, the large outflowing velocity may provide hints on its launch radius. 
To illustrate this, we express the motion equation of the radiatively-driven outflow as
\begin{equation}
\frac{v dv}{dR} = \frac{f_L L}{4 \pi R^2 c m_{\rm p} \nh} - \frac{G M_{\rm BH} }{R^2},
\label{eq:motion}
\end{equation}
where $v$ is the radial wind velocity, $R$ is the radial distance, $f_L$ is the fraction of continuum being absorbed or scattered in the wind,
$L$ is the bolometric luminosity, $c$ is the speed of light, $m_{\rm p}$ is the mass of proton, $M_{\rm BH}$ is the black hole mass, respectively \citep{Hamann1998}. 

Integrating Eq.~\ref{eq:motion} from the launch radius to infinity yields the terminal velocity
\begin{equation}
v_\infty \approx 3200\, R_l^{-1/2} (\frac{f_{0.1} L_{46}}{N_{22}} - 0.084\, M_8)^{1/2},
\label{eq:velocity}
\end{equation}
where $R_l$ is the launch radius in units of pc, $f_{0.1}$ is $f_L/0.1$, $N_{22}$ is the column density relative to $10^{22}\ \nhu$, $L_{46}$ is the bolometric luminosity relative to $10^{46}\ \ergs$, and $M_8$ is the black hole mass in units of $10^8\ M_\odot$, respectively.

The black hole mass was estimated to be $4.1 \times 10^6\ M_\odot$ with a 0.43 dex uncertainty in L15. 
Using the observed maximum velocity as a proxy for terminal velocity and assuming $N_{22} / f_{0.1} \approx 2$ \citep{Leighly2009}, we obtain an average launch radius about $8 \times 10^{-4}$ pc if the winds are produced in the dust-free region, consistent with \cite{Leighly2009}. 
Note that the value of $N_{22} / f_{0.1}$ used here was calculated in \cite{Leighly2009} by approximately reconstructing the intrinsic X-ray to optical SED for WPVS 007 with the use of Mrk 335 and Mrk 493 SEDs as unattenuated templates. 
It is impossible to accurately measure this ratio during the 2010 to 2017 \hst~observations owing to the limited \hst~spectral coverage and weak X-ray signal. But we note that, adopting a larger (e.g., $N_{22} / f_{0.1} = 3$) or smaller (e.g., $N_{22} / f_{0.1} = 1$) ratio would obtain a launch radius about $4 \times 10^{-4}\rm \ pc$ or $2 \times 10^{-3}\rm \ pc$, which will not significantly influence our results.

We then calculate the evolution of wind velocity as a function of acceleration time and radial distance with different initial conditions (\lbol, $R_l$, and $N_{22}/f_{0.1}$) using Eq.~\ref{eq:motion}. 
The results are displayed in Figure \ref{fig:velocity}. For a launch radius as small as $\sim 8 \times 10^{-4}$ pc, the outflow soon reaches its maximum velocity in a few tens of days, and an increase in velocity with luminosity can be clearly seen, which may also contribute to the coordination between outflow velocity and AGN luminosity, as discussed in Section \ref{subsec:velocity_lumin}. 

According to the top panel of Figure \ref{fig:velocity}, the outflow velocity reaches $\rm \sim 10000\ km\ s^{-1}$ at a radial distance $\rm \sim 0.01\ pc$, which is close to the torus sublimation radius estimated in Section \ref{subsec:swift}.
Therefore, the strongly blueshifted BAL features observed may be located far away from their launch radius and close to the torus, since the winds propagate and are accelerated outward (see Figure \ref{fig:velocity}).

To test whether the outflows can be initially launched from the torus scale \citep{Leighly2015}, we assume that the torus can fully absorb the continuum (i.e., $f_L$ = 1.0) to simulate the boosted radiation pressure caused by the presence of dust since dust has a much larger cross section compared with pure electron scattering. We adopt two column densities, i.e., the maximum ($\nh = 1.7 \times \cm$) and minimum ($\nh = 2.8 \times 10^{22}\ \nhu$) column densities reported in \cite{Grupe2013} during 2005--2013 \swift~observations, to calculate the launch radius. 
The derived $R_l$ values are $\rm 3 \times 10^{-3}\ pc$ and $\rm 3 \times 10^{-4}\ pc$, respectively, which are all much smaller than the torus sublimation radius, thereby conflicting with our initial assumption that the outflows are launched in the torus. In addition, the fully absorbed scenario is also inconsistent with observations \citep{Leighly2009}; and more sophisticated calculations of radiation pressure implementing on dusty gas from the torus also indicated that the typical maximum outflow velocity is a few thousand km s$^{-1}$ and hardly reaches $v\rm > 10000\ km\ s^{-1}$, even if an enhanced dust to gas ratio is adopted for luminous dusty quasars \cite[e.g.,][]{Ishibashi2016, Ishibashi2017}. Therefore, we argue that the outflows in WPVS 007 may most probable to be launched in the region closer to the black hole, and the torus origin maybe less likely. But we note that since we only consider radiation pressure in our simple calculation, it is currently unclear how additional factors, such as the existence of magnetic field, would influence our result. Future numerical simulations are required to test whether outflows launched at a scale as large as the torus can reach the observed maximum velocity.

\subsection{Comparison with previous analysis}
\label{subsec:cmp}
L15 captured an occultation event lasting for $\sim 60$ days characterized by a continuous dimming of \uvm~flux accompanied with the increase of $E(B-V)$. Through extensive color and timescale analyses, they concluded that such an event can be explained by variable extinction when different parts of a rotating torus moving into/out of our sightline. In addition, to explain the decreased BAL velocity during the occultation event, L15 proposed that the dusty gas is ablated from the rotating torus with variable scale height; and, the velocity of the outflowing gas decreases with increasing scale height (see Figure 5 and Section 3.4 in L15).  Therefore, as the scale height increases during the occultation event, the torus may intercept our sightline, absorb the continuum flux and reduce the blue-shifted BAL velocity. As \wpvs~emerges from the occultation event, a bluer and brighter SED with faster outflows will be revealed along our sightline again; this prediction is indeed consistent with the new \hst~observation in 2017.

However, we note that if outflows are launched close to the inner accretion disk as we conclude in Section \ref{subsec:torus}, whether the variable-height torus rotates or not should not significantly affect the fast outflows, thus BALs.  Alternatively, since we have already shown that the luminosity for this source must vary significantly, the UV-optical-MIR flux variabilities (including the occultation event, whose light curves can also be modeled by the AGN stochastic variability; see the lower-left panel of Figure 1), the coordinated BAL variabilities (both velocity and strength) with luminosities and the large outflow velocities can all be easily explained, if the variability is driven by the luminosity variations of the central engine and the outflows are launched close to the inner accretion disk.

\section{Conclusion}
\label{sec:conclusion}
WPVS 007, a low-luminosity NLS1, is noted for its unique spectral variability: 
being an X-ray bright, extremely soft AGN transforming into an X-ray weak source, 
the emergence of strong BALs compared with earlier observations, 
and dramatic outflows with $V_{\rm max}$ up to $14000$~$\rm km\ s^{-1}$ given its low luminosity.
Therefore, this source can serve as an ideal laboratory to investigate the physical conditions of the extraordinary low-luminosity, NLS1-BAL systems.

To this end, we have intensively analyzed all the archival \wise, \swift, and \hst~observations of WPVS 007 in order to understand its peculiar variability nature and put important constraints on the driving mechanism of variable fast outflows in this rare object. 

The observed significant MIR variability, the UV-optical color variabilities in the \swift~bands that deviate from the prediction of pure dust attenuation models, and the fact that \swift~light curves can be well fitted by the stochastic AGN variability model suggest that, 
the variability of intrinsic luminosity caused by the change of the inner accretion disk, rather than variable dust extinction, 
should be mainly responsible for these observed facts in \wpvs. 

We also test whether the variability behavior of BALs is consistent with being photoionization-driven and constrain the outflow launch radius. 
The ionization state of \wpvs, traced by its weak \nv~line, might be much lower compared with other luminous BAL quasars, 
possibly related to its low luminosity and weak X-ray nature. 
At the low-ionization state, the \civ~and \siv~column densities and luminosities are positively correlated. 
Owing to the reduction of optical depth as the \wpvs~luminosity decreases, the \civ~and \siv~BAL troughs become weaker; the high-velocity components of the outflow that typically have smaller absorption strengths might be too weak to be detected, resulting in the coordination between the ``observed'' maximum velocity and its bolometric luminosity. 
Based on the radiatively-driven wind model, we estimate that the outflow launch radius might be as small as $\sim 8 \times 10^{-4} \rm\ pc$. However, a large-scale origin (e.g., the torus) cannot be fully excluded owing to the unknown effects from additional factors, e.g., magnetic field; future numerical simulations are required to test whether outflows launched at the torus scale can produce the observed maximum velocity.

\section{ACKNOWLEDGMENTS}
We thank Guilin Liu, Luming Sun and Guobin Mou for useful discussions.
J.Y.L., M.Y.S., and Y.Q.X. acknowledge support from the NSFC (NSFC-11603022, NSFC-11473026, NSFC-11421303, NSFC-11890693), the 973 Program (2015CB857004), the CAS Frontier Science Key Research Program (QYZDJ-SSW-SLH006) and the K.C.Wong Education Foundation. 
M.Y.S. acknowledges support from the China Postdoctoral Science Foundation (2016M600485). T.G.W. and Z.C.H. acknowledge support from the 973 Program (2015CB857005) and the NSFC (NSFC116203021,
NSFC-11703022, NSFC-11833007). 

This publication makes use of data products from the \textit{Wide-field Infrared Survey Explorer}, which is a joint project of the University of California, Los Angeles, and the Jet Propulsion Laboratory/California Institute of Technology, funded by the National Aeronautics and Space Administration. This publication makes use of data products from the \textit{Near-Earth Object Wide-field Infrared Survey Explorer} (NEOWISE), which is a project of the Jet Propulsion Laboratory/California Institute of Technology. NEOWISE is funded by the National Aeronautics and Space Administration. This publication is based on observations made with the NASA/ESA Hubble Space Telescope, obtained from the data archive at the Space Telescope Science Institute. STScI is operated by the Association of Universities for Research in Astronomy, Inc. under NASA contract NAS 5-26555. We acknowledge the use of the public data from the Swift data archive and the UK Swift Science Data Center.

\bibliographystyle{mnras}
\input{cite.bbl}

\bsp	
\label{lastpage}
\end{document}